\documentclass[12pt]{iopart}
\usepackage{graphicx}
\usepackage{cite}
\usepackage{amsfonts}
\usepackage{array}
\usepackage{booktabs}
\usepackage{srcltx}
\usepackage{epsfig}

\begin{document}
\title[Exactly solvable models and ultracold Fermi gases]
      {Exactly solvable models and ultracold Fermi gases}
\author{M. T. Batchelor$^{1,2}$, A. Foerster$^3$\footnote{Corresponding author. Email:angela@if.ufrgs.br}, 
X.-W. Guan$^1$ and C. C. N. Kuhn$^3$}

\address{$^1$  Department of
Theoretical Physics, Research School of Physics and Engineering,\\
Australian National University, Canberra ACT 0200, Australia\\}
\address{$^2$  Mathematical Sciences Institute, Australian National University,\\ 
Canberra ACT 0200, Australia\\}
\address{$^3$ Instituto de F\'{\i}sica da UFRGS, 
Av. Bento Gon\c{c}alves 9500, Porto Alegre, RS, Brazil}

\begin{abstract}
Exactly solvable models of ultracold Fermi gases are reviewed via their thermodynamic Bethe Ansatz solution. 
Analytical and numerical results are obtained for the thermodynamics and ground state properties 
of  two- and three-component one-dimensional attractive fermions with population imbalance. 
New results for the universal finite temperature corrections are given for the two-component model.
For the three-component model, numerical solution of the dressed energy equations 
confirm that the analytical expressions for the critical fields and the resulting phase diagrams 
at zero temperature are highly accurate in the strong coupling regime.
The results provide a precise description of the quantum phases and universal 
thermodynamics which are applicable to experiments with cold  fermionic atoms confined to one-dimensional tubes.
\end{abstract}


\pacs{02.30.Ik, 03.75.Ss, 03.75.Hh, 64.70.Tg}
                     
\parskip 2mm 
\vglue 5mm

\section{Introduction}

The experimental realization of Bose-Einstein condensates in 
dilute atomic gases \cite{Anderson,Bradley,Davies} led to an explosion 
of ongoing research in ultracold matter physics. 
A remarkable development  was the creation of the first 
molecular condensate in an ultracold degenerate Fermi gas\cite{jin1,jin2}, after which 
the condensation of fermionic pairs was soon detected and shown to be a superfluid \cite{ket}.
A relevant question in this fermionic context is 
if superfluidity can persist in a Fermi gas with imbalanced spin population.
In principle, superfluidity may still occur in a mismatched case, and 
some theories with unusual pairings and exotic phases have been proposed, such as 
the Fulde-Ferrell-Larkin-Ovchinnikov (FFLO) phase, among others \cite{FFLO,MF-FFLO}.
Subsequently the search for experimental confirmation of these new phases
has been conducted by different groups \cite{ket2,hulet1}. Until now, however,  
only paired and polarized phases have been detected in three dimensions.

Further remarkable developments have involved the optical confinement of ultracold atoms 
to one dimension (1D), whereby atoms are trapped and cooled in an array of 1D tubes. 
These systems include bosonic Rb atoms \cite{weiss,g2}  and fermionic $^{40}$K atoms \cite{1D-F}.  
The most recent experimental breakthroughs involve 1D Rb atoms in the attractive regime \cite{STG} 
and the realization of a 1D spin-imbalanced attractive Fermi gas of $^6$Li atoms under the degenerate
temperature \cite{Hulet}.
This experimental work in 1D has highlighted the fundamental nature of exactly solvable models of 
quantum many-body systems.
In particular, the Bethe Ansatz (BA) integrable models of Lieb and Liniger \cite{Lieb} for spinless bosons and of 
Yang \cite{Yang} and Gaudin \cite{Gaudin} for two-component fermions.
The experimental work has further highlighted the deep and enduring significance of the BA \cite{Batchelor}.
More general BA integrable multi-component fermions were initially studied by Sutherland \cite{Sutherland}.
In this paper we examine the BA integrable two- and three-component attractive 1D Fermi gases with population imbalance.
We shall see that these ultracold Fermi gases may be used to create nontrivial and exotic phases of matter.
They also pave the way for the direct observation and further study of FFLO-like states.

\section{Two-component attractive Fermi gas with polarization}

\subsection{The model}

We begin by reviewing the exactly solved two-component model, with  Hamiltonian  \cite{Yang,Gaudin} 
\begin{eqnarray}
{\cal H} = -\frac{\hbar^2}{2m}\sum_{i=1}^{N}\frac{\partial^2}{\partial x_{i}^2} + g_{1D}
\sum_{1\leq i<j\leq N} 
\delta(x_i - x_j) - \frac12{H}(N_{\uparrow}-N_{\downarrow})
\label{Ham}
\end{eqnarray}
which describes $N$ $\delta$-interacting spin-$\frac{1}{2}$ fermions of
mass $m$ constrained
by periodic boundary conditions to a line of length $L$ and subject to an external magnetic field $H$.
The inter-component interaction $g_{\rm 1D}$ can be tuned from strongly attractive
to strongly repulsive via Feshbach resonance and optical
confinement. The interaction is attractive for $g_{\rm 1D}<0$ and
repulsive for $g_{\rm 1D}>0$. Here we will focus on the 
strongly attractive case, since this regime can be experimentally reached 
in 1D \cite{Hulet}.
For convenience, we define $c = mg_{1D}/\hbar ^2$ and a dimensionless interaction strength
$\gamma=c/n$ for the physical analysis, with linear density $n ={N}/{L}$.

The model (\ref{Ham}), which exhibits $SU(2)$ symmetry, was solved  independently by Yang \cite{Yang}   
and Gaudin \cite{Gaudin} using the nested BA.  
The energy eigenspectrum is given by
\begin{equation}
E=\frac{\hbar ^2}{2m}\sum_{j=1}^Nk_j^2, 
\label{energy}
\end{equation}
where  the quasimomenta $\left\{k_j\right\}$ of the fermions, satisfy the BA equations \cite{Yang,Gaudin}
\begin{eqnarray}
&& \exp(\mathrm{i}k_jL) = \prod^M_{\ell = 1}
\frac{k_j-\Lambda_\ell+\mathrm{i}\, c/2}{k_j-\Lambda_\ell-\mathrm{i}\, c/2},\nonumber\\
&& \prod^N_{\ell = 1}\frac{\Lambda_{\alpha}-k_{\ell}+\mathrm{i}\, c/2}{\Lambda_{\alpha}-k_{\ell}-\mathrm{i}\, c/2}
 = - {\prod^M_{ \beta = 1} }
\frac{\Lambda_{\alpha}-\Lambda_{\beta} +\mathrm{i}\, c}{\Lambda_{\alpha}-\Lambda_{\beta} -\mathrm{i}\, c}, 
\label{BEtwo}
\end{eqnarray}
for  $j = 1,\ldots, N$ and $\alpha = 1,\ldots, M$, with $M$ the number of spin-down fermions. 
Here $\left\{\Lambda_{\alpha}\right\}$ are the rapidities for the
internal spin degrees of freedom.

The solutions to the BA equations (\ref{BEtwo}) provide 
the ground state properties and the elementary excitations of the model.
It was shown \cite{Takahashi,BBGO} that the distribution of the quasimomenta in the complex plane for 
the ground state involves bound states, which can be interpreted as BCS-like pairs, 
and unpaired (excess) fermions.   
In the thermodynamic limit, i.e., $N,L \to \infty$ with $N/L$ finite, 
and at zero temperature, all quasimomenta $k_j$ of $N$ atoms form two-body bound states,
i.e., $k_j=\Lambda_j \pm   \mathrm{i} \frac{1}{2} c$ for $j=1,\ldots,M$, accompanied by the real spin parameter
 $\Lambda_j$\cite{Takahashi}. 
The BA equations (\ref{BEtwo}) then become
\begin{eqnarray}
k_j L& =& 2\pi I_j + \sum_{l=1}^{M}
\theta \left(\frac{k_j-\Lambda_l}{c}\right),\qquad j=2M+1,\ldots, N\\
2 \Lambda_j L &=& 2 \pi J_j +
\sum_{l=1}^{N-2M}\theta \left(\frac{\Lambda_j-k_l}{c} \right)+ \sum_{l=1}^{M}
\theta \left(\frac{\Lambda_j-\Lambda_l}{2 c} \right),\, \, \,
j=1,\ldots, M.\nonumber
\end{eqnarray}
where $\theta(x) = 2 \arctan 2 x$ with $I_j=-(N-2M-1)/2,-(N-2M-3)/2,\ldots,(N-2M-1)/2$ and $J_j=-(M-1)/2,\ldots, (M-3)/2,(M-1)/2$.

Introducing the density of unpaired fermions $\rho(k)=dI_j(k)/Ldk$ and the density of 
pairs $\sigma(\Lambda)=dJ_j(\Lambda)/Ld\Lambda$ it follows that 
\begin{eqnarray}
\rho(k)&=&\frac{1}{2\pi}-\frac{1}{2\pi}\int_{-B}^B\frac{|c|\sigma(\Lambda)d\Lambda}{c^2/4+(k-\Lambda)^2},\nonumber\\
\sigma(\Lambda)&=&\frac{1}{\pi}-\frac{1}{2\pi}\int_{-B}^B\frac{2|c|\sigma(\Lambda')d\Lambda'}{c^2+(\Lambda-\Lambda')^2}
-\frac{1}{2\pi}\int_{-Q}^Q\frac{|c|\rho(k)dk}{c^2/4+(\Lambda-k)^2} \label{BE-string}
\end{eqnarray}
in terms of which the total number of particles, the magnetization and the ground state energy per unit length are given by
\begin{eqnarray}
n=2\int_{-B}^B\sigma(\Lambda)d\Lambda+\int^Q_{-Q}\rho(k)dk,\\
M^z=\frac{1}{2}(n_{\uparrow}-n_{\downarrow})=\frac{1}{2}\int_{-Q}^Q\rho(k)dk, \\
\frac{E}{L}=\int_{-B}^{B}\left(2\Lambda^2-c^2/2\right)\sigma(\Lambda)d\Lambda+\int_{-Q}^Qk^2\rho(k)dk.
\end{eqnarray}
Here the integration boundary  $Q$ characterises  the Fermi momentum in quasi-momentum space and $B$ is the Fermi momentum  in spin rapidity space. 

\begin{figure}
\begin{center}
\includegraphics[width=12cm,angle=0]{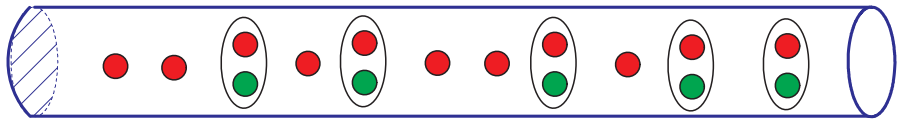}
\caption{Schematic representation showing the mixture of bound pairs and excess fermions in the strongly attractive regime.}
\label{fig:LL}
\end{center} 
\end{figure}

The BA equations (\ref{BE-string}) are  essential to understanding the ground state properties of the model.
The integration boundaries implicitly depend on the total number of particles and the magnetization.  
The strong coupling expansion for the ground state energy per unit length 
\begin{eqnarray}
\frac{E}{L} &\approx& \frac{\hbar^2n^3}{2m}\left\{ -\frac{(1-P)\gamma^2}{4} + \frac{P^3\pi^2}{3}\left[ 1 + \frac{4(1-P)}{|\gamma|}\right]\right. \nonumber\\
&&\left. +\frac{\pi^2(1-P)^3}{48}\left[1+\frac{(1-P)}{|\gamma|}+\frac{4P}{|\gamma|}\right] \right\}
\label{Estrong}
\end{eqnarray}
has been obtained by using the discrete BA equations (\ref{BEtwo}) \cite{BBGO} or by iteration with the integral equations \cite{Wadati}.
Here the polarization $P$ is defined by $P=(N_{\uparrow}-N_{\downarrow})/N$ and the      
binding energy is $\epsilon_b=\hbar^2n^2\gamma^2/4m$. 
The explicit relation between the polarization and the external field $H$ is given further below in equation (\ref{Hstrong}). 
In this strong coupling region, the system behaves like a mixture of bound pair composites 
and excess single particles. The form of the ground state energy (\ref{Estrong}) reveals the 
nature of weak pair-pair attractive interaction and pair-unpair attractive 
interaction.  A schematic representation of this mixture is depicted in Figure  \ref{fig:LL}.

\subsection{Thermodynamic Bethe Ansatz}

The thermodynamic properties as well as quantum phase transitions in this model system can be analyzed 
through the Thermodynamic Bethe Ansatz (TBA). This method was originally proposed by Yang and Yang \cite{YangYang} 
for spinless bosons and received important contributions for fermions by Takahashi \cite{Takahashi-T}, among others \cite{Takahashi-B}.
The basic idea of the method is that in the
thermodynamic limit we can consider a continuous distribution function for the BA roots. 
The equilibrium state can be obtained by the condition of minimizing the Gibbs free energy $G$, 
\begin{equation}
G = E - HM^z - \mu N - TS, 
\end{equation}
where $\mu$ is the chemical potential and $S$ the entropy.

This procedure gives rise to a set of coupled nonlinear equations, the TBA equations, for 
the dressed energies $\epsilon^{\rm b}$ for paired and $\epsilon^{\rm u}$ for unpaired fermions 
(see \cite{Takahashi-T,Takahashi-B,GBLB} for details)
\begin{eqnarray}
&&\epsilon^{\rm b}(k)=2(k^2-\mu-\frac14{c^2})+Ta_2*\ln(1+\mathrm{e}^{-\epsilon^{\rm b}(k)/T} )
+ \, Ta_1*\ln(1+\mathrm{e}^{-\epsilon^{\rm u}(k)/{T}}), \nonumber\\
&& \epsilon^{\rm u}(k)=k^2-\mu-\frac12{H}+Ta_1*\ln(1+\mathrm{e}^{-\epsilon^{\rm  b}(k)/{T}})
-T\sum_{m=1}^{\infty}a_n*\ln(1+\eta_n^{-1}(k)), \nonumber\\
&& \ln \eta_n(\lambda)=\frac{nH}{T}+a_n*\ln(1+\mathrm{e}^{-\epsilon^{\rm u}(\lambda)/{T}})
+\sum_{m=1}^{\infty}T_{nm}*\ln(1+\eta^{-1}_m(\lambda)) \label{TBA}
\end{eqnarray}    
where  $*$ denotes the convolution integral $(f*g)(\lambda)=\int^{\infty}_{-\infty}f(\lambda-\lambda')g(\lambda')d\lambda'$ 
and the function $\eta_n$ is the ratio of string densities, coming
from Takahashi's hypothesis that complex quasimomenta group together to form strings in the complex plane \cite{Takahashi-B}.
The functions 
\begin{equation}
a_m(\lambda)=\frac{1}{2\pi}\frac{m|c|}{(m \,c/2)^2+\lambda^2} 
\end{equation}
and $T_{mn}(\lambda)$ can also be found in Takahashi's book \cite{Takahashi-B}.

In the limit when $T\to 0$ the TBA equations reduce to the dressed energy equations 
\begin{eqnarray}
 {\epsilon^{\rm b}}(\Lambda)=2(\Lambda^2-\mu -\frac{c^2}{4})
 -\int_{-B}^{B} a_2(\Lambda-\Lambda'){\epsilon^{\rm b}}(\Lambda')d\Lambda'
 -\int_{-Q}^{Q} a_1(\Lambda-k){\epsilon^{\rm u}}(k)d k, \nonumber\\ 
{\epsilon^{\rm u}}(k)=(k^2-\mu-\frac{H}{2})-\int_{-B}^{B}a_1(k-\Lambda){\epsilon^{\rm b}}(\Lambda)d\Lambda.
\label{epstwo}
\end{eqnarray}
The Gibbs free energy per unit length at zero temperature can be written in terms of the dressed energies as
\begin{equation}   
G(\mu,H)={\frac{1}{\pi}\int_{-B}^B{\epsilon^{\rm b}}(\Lambda )d\Lambda}+\frac{1}{2\pi}\int_{-Q}^Q{\epsilon^{\rm u}}(k)dk, 
\label{gibbs}      
\end{equation}
from which the density of fermions and the magnetization (or polarization $P$) follow via the relations
\begin{equation}
 -\partial G(\mu,H)/\partial \mu =n, \,\,\,\,-\partial
G(\mu,H)/\partial H = m_z = nP/2. \nonumber
\label{part}
\end{equation}
In general, the dressed energy equations
provide a clear picture of band fillings with respect to the field
$H$ and the chemical potential $\mu$ at arbitrary temperatures \cite{BGOS}. 
These equations can be analytically solved in some special limits, such as in the strongly attractive
regime $|\gamma| \gg 1$ \cite{GBLB} discussed below. For comparison, we also present their numerical solution \cite{jingsong}.

\subsection{Phase diagram at zero temperature}

The set of equations (\ref{epstwo}), (\ref{gibbs}) and (\ref{part}) were solved in \cite{GBLB,jingsong} 
for strongly attractive interaction using a lengthy iteration method. 
To leading order, the explicit form for the external field in terms of 
the density of fermions, the polarization and the interaction strength is given by
\begin{eqnarray}
H & = & \frac{\hbar^2n^2}{2m}\left\lbrace \frac{\gamma^2}{2} + 
2P^2\pi^2\left[1+\frac{4(1-P)}{|\gamma|} - \frac{4P}{3|\gamma|}\right]  \right. \nonumber\\
&& \left. \qquad \qquad - \frac{\pi^2(1-P)^2}{8}\left[1+\frac{4P}{|\gamma|}\right]\right\rbrace. \qquad
\label{Hstrong}
\end{eqnarray}
This equation determines the full phase diagram of the model and the critical values of the external field 
\begin{eqnarray}
H_{c1} &\approx& \frac{\hbar^2n^2}{2m}\left( \frac{\gamma^2}{2} - \frac{\pi^2}{8}\right), \nonumber\\
H_{c2} &\approx& \frac{\hbar^2n^2}{2m}\left[ \frac{\gamma^2}{2} + 2\pi^2\left(1-\frac{4}{3|\gamma|}\right)\right].
\label{hcritico}
\end{eqnarray}
These results for $H_{c1}$ and $H_{c2}$ are obtained from equation (\ref{Hstrong}) by setting the 
polarization to $P=0$ and $P=1$, respectively. The critical fields can be physically interpreted as 
separating three distinct quantum phases: (i) for $H<H_{c1}$ bound pairs populate the ground state, 
(ii) for $H>H_{c2}$ a completely ferromagnetic phase occurs, and (iii) 
for the intermediate range $H_{c1}<H<H_{c2}$ paired and unpaired atoms coexist.

\begin{figure}[t]
{\includegraphics[width=0.80\linewidth,angle=-90]{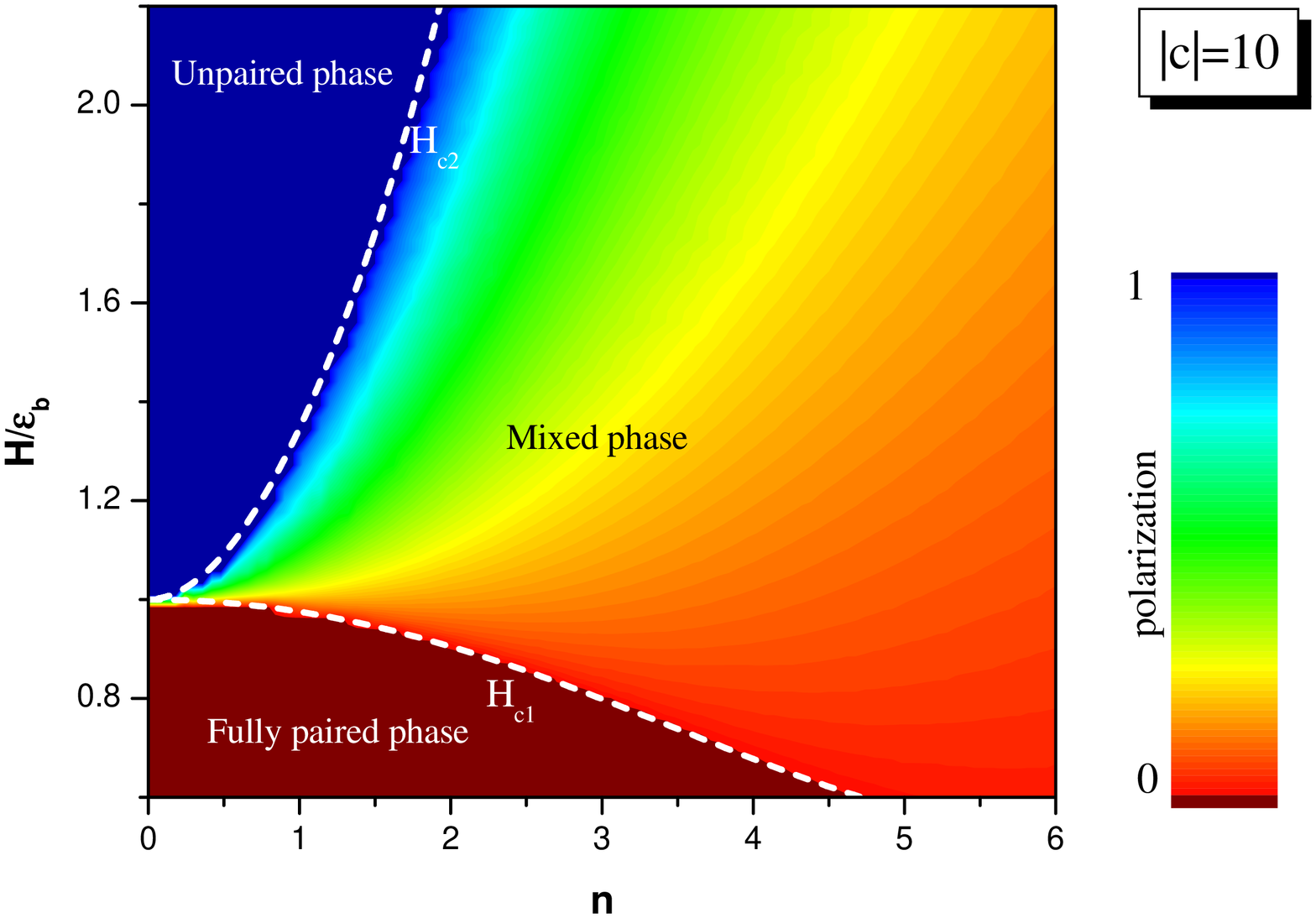}}
\caption{Phase diagram for strong coupling $|c| = 10$ in the $n-H$ plane (from Ref.~\cite{jingsong}).
Good agreement is found between the analytical results (\ref {hcritico}) for the critical fields 
represented by the dashed lines 
and the numerical solution of the dressed energy equations (\ref{epstwo}), given
by the coloured phases.}
\label{nstrong}
\end{figure}

Figure \ref{nstrong} illustrates the phase diagram in the $n-H$ plane for the particular value $|c| = 10$.
The  dashed lines are plotted from equations (\ref{hcritico}). The coloured phases
are obtained by numerically integrating the dressed energy equations (\ref{epstwo}).
The analytical results coincide well with the numerical boundaries.
Here it is also clear that subject to the value of the external field, 
the system exhibits three quantum phases: 
a BCS-like fully paired phase; an unpaired, fully polarized phase; 
and a mixed, partially polarized phase, composed of BCS-like pairs and unpaired (excess) fermions.
The mixed phase can be considered as a 1D-analogue of the FFLO phase (see, e.g. \cite{Parish,Liu,cazalilla2,Batrouni}).

Calculations on the BA integrable model thus lead to 
a two-shell structure composed of the partially polarized FFLO-type phase in the 
centre of the trap surrounded by -- depending on the strength of 
the external field -- either fully paired or fully polarized wings \cite{Orso, Hu, GBLB}.
This prediction was recently verified by the observation by Liao {\em et al.} \cite{Hulet} of three distinct phases 
in experimental measurements of ultracold ${}^{6}$Li atoms in an array of 1D tubes. 
A number of other authors have also investigated the BA integrable model (\ref{Ham}) in the context of the 
Fermi gas (see, e.g., \cite{Wadati1a,Wadati2,kaka,ZGLBO}).

\subsection{TBA for low temperature}

In order to handle with the TBA equations (\ref{TBA}) at low temperature, we can employ an expansion in terms of the polylogarithm 
function, which has so far been applied to the 1D attractive Fermi gases of ultracold atoms up to order $1/c^2$ \cite{ZGLBO,sun,he}. 
This approach is widely applicable to 1D many-body systems with quadratic or linear bare dispersions 
in both the attractive and repulsive regimes.  
It is also expected that the analytical polylogarithmic function approach will play a central role in
unifying the properties of attractive Fermi gases of ultracold atoms with higher symmetries. 
For example, the TBA equations for the 1D Fermi gases with $\delta$-function attractive interactions and internal
spin degrees of freedom may be reformulated according to the
charge bound states and spin strings characterizing spin fluctuations \cite{sun}. 
Thus for strong attraction, the spin fluctuations that couple to non spin-neutral charge bound states are exponentially small
and can be asymptotically calculated \cite{sun}. 
Thus the low energy physics is dominated by density fluctuations among the charge bound states. 
The full phase diagrams and thermodynamics of the 1D attractive Fermi gases can be analytically calculated via this
polylogarithm function approach. 
For spin-1/2 attractive fermions, universal Tomonaga-Luttinger liquid behaviour
was identified from the pressure given in terms of polylog functions \cite{ZGLBO}. 
Here we discuss some further results, on the universal nature of finite temperature corrections.

For $H \gg k_BT$, where $k_B$ is the Boltzmann constant, we can  ignore  the exponentially small corrections from the 
spin bound states and
collect the terms up to $1/\gamma^2$ in the TBA equations (\ref{TBA}), resulting in the simple form 
\begin{eqnarray}
\epsilon^{\rm b}(k) \approx  2k^2-A^{\rm b}(T,H), \nonumber \\ 
\epsilon^{\rm u}(k)\approx k^2-A^{\rm u}(T,H), \label{TBA2}
\end{eqnarray}
where
\begin{eqnarray}
A^{\rm b}(T,H) &=& 2\left[\mu+\frac14{c^2}-\frac{1}{2|c|}\left(p^{\rm
  b}(T,H)+4p^{\rm u}(T,H)\right)\right] + O\left(\frac{1}{c^3}\right),\\
A^{\rm u}(T,H)&=&\mu+T\ln(2\cosh(\frac{H}{2T}))-\frac{2}{|c|}p^{\rm
  b}(T,H)\nonumber\\
&& -\frac{J}{4}\left(1-B^2 \right)+\frac{3J^2}{32k_BT}(1-B^4) + O\left(\frac{1}{c^3}\right). \label{A_u}
\end{eqnarray}
Here we have denoted $B=\tanh\frac{H}{2T}$ and $J=\frac{2}{|c|}p^{\rm u}(T,H)$.  
In order to examine the low temperature Luttinger liquid signature, we can safely ignore
 the exponentially small contribution from the spin wave bound states in the function $A^{\rm u} $.  
Thus integration by parts gives  the effective pressures 
\begin{eqnarray}
p^{\rm b}(T,H)&=&\frac{\sqrt{2}}{\sqrt{\frac{\pi^2\hbar^2}{2m}}}\int_0^{\infty}\frac{\sqrt{\epsilon} \, d\epsilon
}{1+\mathrm{e}^{\frac{\epsilon-A^{\rm b}(T,H)}{k_BT}}},\nonumber\\
p^{\rm u}(T,H)&=&\frac{1}{\sqrt{\frac{\pi^2\hbar^2}{2m}}}\int_0^{\infty}\frac{\sqrt{\epsilon} \, d\epsilon
}{1+\mathrm{e}^{\frac{\epsilon-A^{\rm u}(T,H)}{k_BT}}}.\label{P}
\end{eqnarray}
The integrals in (\ref{P}) can be calculated explicitly using
Sommerfeld  expansion at low temperatures. We assume that there exist two Fermi seas, 
i.e., a Fermi sea of bound pairs with a cut-off potential $A^{\rm  b}(T,H)/2$ and a Fermi sea of 
unpaired fermions with a cut-off potential $A^{\rm u}(T,H)$.  
After some lengthy iteration with the relations (\ref{part}) we obtain the free energy  
\begin{eqnarray}
F(T,H) &\approx&
\frac{\hbar^2\pi^2n^3P^3}{6m}\left[ 1-
\frac{\pi^2}{4}\left(\frac{k_BT}{\mu_0^{\rm
    u}}\right)^2-\frac{\pi^4}{60}\left(\frac{k_BT}{\mu_0^{\rm
    u}}\right)^4\right.\nonumber\\
&&\left. +\frac{4(1-P)}{|\gamma|}\left(1+\frac{\pi^2}{4}\left(\frac{k_BT}{\mu_0^{\rm
    u}}\right)^2+\frac{2\pi^4}{15}\left(\frac{k_BT}{\mu_0^{\rm
    u}}\right)^4 \right)\right] \nonumber\\
&&+\frac{\hbar^2\pi^2n^3}{2m}\frac{(1-P)^3}{48}\left[ 1-
\frac{\pi^2}{16}\left(\frac{k_BT}{\mu_0^{\rm
    b}}\right)^2-\frac{\pi^4}{960}\left(\frac{k_BT}{\mu_0^{\rm
    b}}\right)^4\right.\nonumber\\
&&\left. +\left(\frac{(1-P)}{|\gamma|}+\frac{4P}{|\gamma|}\right)\left(1+\frac{\pi^2}{16}\left(\frac{k_BT}{\mu_0^{\rm
    b}}\right)^2+\frac{\pi^4}{120}\left(\frac{k_BT}{\mu_0^{\rm
    b}}\right)^4 \right)\right] \nonumber\\
&&
-\frac{1}{2}nPH-\frac{\hbar^2}{2m}n(1-P)\frac{c^2}{4},\label{FE-T0}
\end{eqnarray}
where $\mu^{\rm u}_0= \frac{\hbar^2\pi^2n^2P^2}{2m}$ and
$\mu^{\rm b}_0=\frac{\hbar^2\pi^2n^2(1-P)^2}{32m}$.
We see clearly that the free energy reduces to the ground state energy (\ref{Estrong}) as $T\to 0$.

For the temperature $k_BT \ll E_F$, where $E_F$ is the Fermi energy, the spin wave fluctuation is frozen out
due to the large magnetic field. The leading low temperature correction to the free energy (\ref{FE-T0}) is
\begin{equation}
F(T,H)=E_0(H)-\frac{\pi  C k_B^2T^2}{6\hbar}\left(\frac{1}{v_{\rm b}}+\frac{1}{v_{\rm u}}\right),\label{FF-C}
\end{equation}
which belongs to the universality class of the Gaussian model with central charge $C=1$ \cite{Affleck}. 
In the above equation, the ground state energy $E_0(H)$ is as given in (\ref{Estrong}), subject to an additional term $-nPH/2$.
The group velocities for bound pairs and unpaired fermions are 
\begin{eqnarray}
v_{\rm b} &\approx &\frac{v_{\rm F}(1-P)}{4}
\left(1+\frac{(1-P)}{|\gamma|} +\frac{4P}{|\gamma|}\right),\nonumber\\
v_{\rm u}&\approx & v_{\rm F} P\left( 1 +
\frac{4(1-P)}{|\gamma|}\right),
\end{eqnarray}
respectively. Here the Fermi velocity is $v_{\rm F}=\hbar \pi n/m$. 
The above result indicates that the low energy physics for 1D strongly
attractive fermions in the gapless phase can be described by a
two-component Tomonaga-Luttinger liquid model as long as the ferromagnetic 
spin-spin interaction is frozen out.

\begin{figure}[ht]
{{\includegraphics [width=0.60\linewidth]{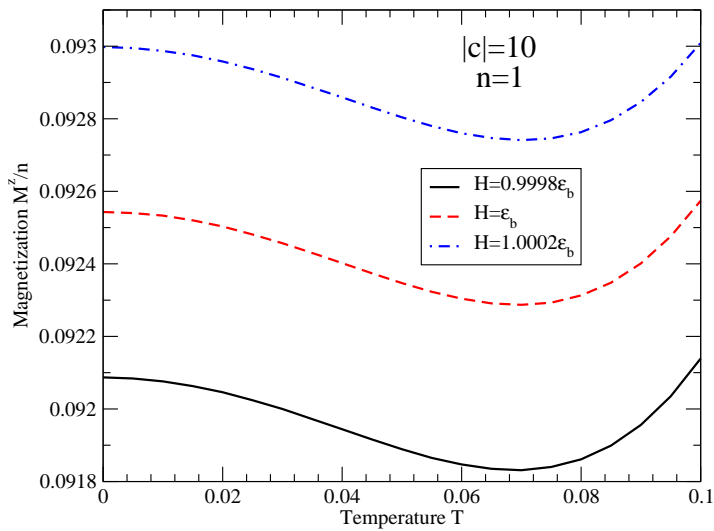}}}
\caption{Magnetization $m^z$ vs temperature at different 
external field values (in the units $2m=\hbar=1$). The minimum of
the magnetization is clearly observed for each field value.} 
\label{fig:mz-m}
\end{figure}

In order to recognize a Tomonaga-Luttinger liquid signature in the gapless
phase in the low temperature limit, we examine the temperature
dependent relations for the magnetization. In general  \cite{Maeda}, one should expect
 a magnetization minimum due to a crossover from a Tomonaga-Luttinger liquid with a 
 linear dispersion to a state governed by the nonrelativistic dispersion $\epsilon\propto k^2$.
This magnetization minimum does exist in the gapless phase in
attractively interacting fermions. We plot the magnetization vs
temperature from the free energy (\ref{FE-T0}) in Figure \ref{fig:mz-m}, where we 
observe a clear minimum of the magnetization for different magnetic fields. 
This comes about due to a crossover from the hardcore bosonic signature of the 
Tomonaga-Luttinger liquid to polarized free fermions.  For further understanding this
signature, we derive the magnetization from (\ref{FF-C}) where we
consider the  low temperature limit, i.e., $T \to 0$ (in natural units)
\begin{equation}
m^z=m^z_0-\frac{\pi T^2}{6}\left(\frac{1}{v_b^2}\frac{\partial
  v_b}{\partial H} +\frac{1}{v_u^2}\frac{\partial
  v_u}{\partial H}\right),\label{mz-t}
\end{equation}
where $m^z_0=\partial E_0(H)/\partial H$. The finite temperature  contribution
to the magnetization at low temperatures depends on the signs from the
term in the brackets  of equation (\ref{mz-t}). 
This part  indicates the existence of a minimum of the magnetization. 
In the gapless phase two Fermi liquids are coupled through pair-unpaired fermion scattering. 
The linear field dependent magnetization is a consequence of the fact that the total number of
fermions is fixed.      

So far we have considered the simplest 1D exactly solvable model in the scenario of ultracold Fermi gases. Generalizations
to three and more components can be performed and are discussed in the next sections.

\section{Three-component attractive Fermi gas with polarization}

\subsection{The model}

To describe a three-component Fermi gas, the same type of Hamiltonian (\ref{Ham}),
with a kinetic and a contact potential interacting term, can be considered.
However, now the 
$N$ fermions can occupy three possible hyperfine
levels ($|1\rangle$, $|2\rangle$ and $|3\rangle$)
and the Zeeman term is expressed in terms of two external
fields. 
The Hamiltonian reads \cite{Sutherland} 
\begin{eqnarray}
{H}&=&-\frac{\hbar ^2}{2m}\sum_{i = 1}^{N}\frac{\partial
^2}{\partial x_i^2}+\,g_{\rm 1D} \sum_{1\leq i<j\leq N} \delta
(x_i-x_j)+\sum_{i=1}^3N^{i}\epsilon^{i}_Z(\mu_B^{i},B). \label{Hamth}
\end{eqnarray}
The last term denotes the Zeeman energy, where $N^{i}$ is the
number of fermions in state $| i\rangle$ with Zeeman energy
$\epsilon^{i}_Z$ determined by the magnetic moments $\mu_B^{i}$
and the magnetic field $B$.
This term can also be written as
$ -H_1(N^1-N^2)-H_2(N^2-N^3) +N\bar{\epsilon}$, where the unequally spaced Zeeman
splitting in three hyperfine levels can be specified by two
independent parameters $H_1 = \bar{\epsilon} -
\epsilon^{1}_Z(\mu_B^{1},B)$ and $H_2 = \epsilon^{3}_Z(\mu_B^{3},B)-
\bar{\epsilon}$, with $\bar{\epsilon}$ the average Zeeman energy.
We use the same notation and conventions as in the previous case.

The Hamiltonian (\ref{Hamth}) exhibits $SU(3)$ symmetry and
was solved by Sutherland in the sixties by means of the nested BA \cite{Sutherland}.
The energy eigenspectrum is again given in terms of the quasimomenta 
$\left\{k_i\right\}$  of the fermions by (\ref{energy}), but now satisfying  \cite{Sutherland,Takahashi}
\begin{eqnarray}
\exp(\mathrm{i}k_jL)=\prod^{M_1}_{\ell = 1} \frac{k_j-\Lambda_\ell+\mathrm{i}\, c/2}{k_j-\Lambda_\ell-\mathrm{i}\, c/2},\nonumber \\ 
\prod^N_{\ell = 1}\frac{\Lambda_{\alpha}-k_{\ell}+\mathrm{i}\, c/2}{\Lambda_{\alpha}-k_{\ell}-\mathrm{i}\, c/2}= - {\prod^{M_1}_{ \beta = 1} } \frac{\Lambda_{\alpha}-\Lambda_{\beta} +\mathrm{i}\, c}{\Lambda_{\alpha}-\Lambda_{\beta} -\mathrm{i}\, c} {\prod^{M_2}_{\ell = 1} }\frac{\Lambda_{\alpha}-\lambda_{\ell} -\mathrm{i}\, c/2}{\Lambda_{\alpha}-\lambda_{\ell} +\mathrm{i}\, c/2},\nonumber \\ 
 {\prod^{M_1}_{ \ell = 1} }\frac{\lambda_{\mu}-\Lambda_{\ell} +\mathrm{i}\, c/2}{\lambda_{\mu}-\Lambda_{\ell} -\mathrm{i}\, c/2}=-{\prod^{M_2}_{\ell = 1} }\frac{\lambda_{\mu}-\lambda_{\ell} +\mathrm{i}\, c}{\lambda_{\mu}-\lambda_{\ell} -\mathrm{i}\, c} 
\label{BEth}
\end{eqnarray}
for $j=1,\ldots, N$, $\alpha = 1,\ldots, M_1$, $\mu=1,\ldots,M_2$, with
quantum numbers $M_1=N_2+2N_3$ and $M_2=N_3$.
The parameters $\left\{\Lambda_{\alpha},\lambda_m\right\}$ are the rapidities for the internal hyperfine spin 
degrees of freedom. For the irreducible representation $\left[3^{N_3}2^{N_2}1^{N_1}
\right]$, a three-column Young tableau encodes the numbers of
unpaired fermions ($N_1=N^1-N^2$), bound pairs ($N_2=N^2-N^3$) and trions ($N_3=N^3$).

The ground state energy per unit length  
\begin{eqnarray}
\frac{E}{L} &\approx& \frac{\pi^2n_1^3}{3}\left(1+\frac{8n_2+4n_3}{|c|}\right) -\frac{n_2c^2}{2}\nonumber\\
& & +\frac{\pi^2n_2^3}{6}\left(1+\frac{12n_1+6n_2+16n_3}{3|c|}\right)-2n_3c^2 \nonumber\\
& & +\frac{\pi^2n_3^3}{9}\left(1+\frac{12n_1+32n_2+18n_3}{9|c|}\right)
\label{E}
\end{eqnarray}
was obtained in \cite{xiwen} by solving the BA equations (\ref{BEth}).
Here $n_a=N_a/L,\,\,a=1,2,3$ are the densities of unpaired fermions, bound pairs and trions, respectively, 
with the constraint $n=n_1 + 2 n_2 + 3 n_3$.

The phase diagram of the system for the strongly attractive interaction regime 
can also be obtained using the dressed energy formalism.
For this model, a richer scenario, with more quantum phases, is expected. 
Indeed, as we shall discuss further below, a trion phase, which consists of three-body bound states, 
along with a number of mixed phases, are found \cite{xiwen}.

\subsection{Dressed energy formalism and phase diagrams at zero temperature}

In the thermodynamic limit the TBA equations are found by minimizing 
the Gibbs free energy $G = E + n_1H_1+n_2H_2-\mu N - TS$.
In the limit $T\to 0$, the dressed energy equations obtained \cite{Takahashi-B,xiwen,Schlot1} are 
\begin{eqnarray}
\epsilon^{(3)}(\lambda)&=&3\lambda^2-2c^2-3\mu-a_2*{\epsilon^{(1)}}(\lambda) \nonumber\\
& &-\left[a_1+a_3\right]*{\epsilon^{ (2)}}(\lambda)-\left[a_2+a_4\right]*{\epsilon^{ (3)}}(\lambda) \nonumber\\
\epsilon^{(2)}(\Lambda)&=&2\Lambda^2-2\mu-\frac{c^2}{2}-H_2-a_1*{\epsilon^{ (1)}}(\Lambda)  \nonumber\\
& &-a_2*{\epsilon^{2}}(\Lambda) -\left[a_1+a_3\right]*{\epsilon^{(3)}}(\Lambda)  \label{TBA-F} \\
\epsilon^{ (1)}(k)&=&k^2-\mu-H_1-a_1*{\epsilon^{ (2)}}(k) -a_2*{\epsilon^{(3)}}(k). \nonumber
\end{eqnarray}
Here $\epsilon^{(a)}$ are the dressed energies
and  $a_j*{\epsilon^{(a)}}(x)=\int_{-Q_a}^{+Q_a}a_j(x-y){\epsilon^{(a)}}(y)dy$.
The negative part of the dressed energies $\epsilon^{(a)}(x)$ for $x\le \left|Q_{a}\right|$ correspond 
to the occupied states in the
Fermi seas of trions, bound pairs and unpaired fermions, with the positive part of 
$\epsilon^{(a)}$ corresponding to the unoccupied states.
The integration boundaries $Q_{a}$ characterize the ``Fermi surfaces'' at $\epsilon^{(a)}(\pm Q_{a})=0$. 
The zero-temperature
Gibbs free energy per unit length can be written in terms of the dressed energies as 
\begin{equation}
G = \sum_{a=1}^3 \frac{a}{2\pi} \int_{-Q_a}^{+Q_a}{\epsilon^{(a)}}(x)dx, 
\end{equation}
from which physical quantities are obtained through the thermodynamic relations
\begin{equation}
-\frac{\partial G}{\partial \mu} =n, \,-\frac{\partial G}{\partial H_1}=n_1,\, -\frac{\partial G}{\partial H_2}=n_2.
\end{equation}

The dressed energy equations (\ref{TBA-F}) can be analytically solved just in some special limits, 
such as the strongly attractive coupling regime.  
This was discussed in \cite{xiwen}, where the analytical expressions 
\begin{eqnarray}
H_1&=&\pi^2n_1^2\left( 1 - \frac{4n_1}{9|c|} + \frac{8n_2}{|c|} + \frac{4n_3}{|c|}\right) + \frac{10\pi^2n^3_2}{27|c|} \nonumber \\
&& - \frac{\pi^2n^2_3}{9}\left(1+\frac{4n_1}{3|c|} + \frac{32n_2}{9|c|} + \frac{4n_3}{3|c|}\right) + \frac{2c^2}{3}, \nonumber \\ 
H_2&=&\frac{\pi^2n_2^2}{2}\left(1+\frac{4n_1}{|c|} + \frac{40n_2}{27|c|} + \frac{16n_3}{3|c|}\right) + \frac{16\pi^2n_1^3}{9|c|} \nonumber \\
&& - \frac{2\pi^2n_3^2}{9}\left(1+\frac{4n_1}{3|c|} + \frac{32n_2}{9|c|} + \frac{8n_3}{9|c|}\right) 
+ \frac{5c^2}{6}, \label{H1-H2}
\end{eqnarray}
for the  energy transfer relations were found (in units of $\hbar^2/2m$). 
These equations determine the full phase diagram and the critical field values activated by the fields $H_{1}$ and $H_{2}$.

Here we numerically solve these equations to confirm the analytical expressions for the physical quantities and
the resulting phase diagrams of the model.
Basically, for each value of the integration boundaries $Q_{a}, a = 1,2,3$, the equations (\ref{TBA-F}) are converted 
into a finite size single matrix equation that can be solved using the condition $\epsilon^{(a)}(\pm Q_{a})=0$ to get 
the dressed energies, the fields and the chemical potential.
Then the Gibbs free energy is obtained and from it the polarizations and the linear density are found.
Each phase can be identified by properly separating the input in distinct sets containing different combinations of 
vanishing/non vanishing integration boundaries.
For example, in the pure unpaired phase we 
have $ \{Q_{2}=Q_{3}=0, Q_{1}\neq 0\}$, in the mixed phased composed of unpaired fermions and bound 
pairs $\{Q_{3}=0, Q_{1}\neq 0, Q_{2}\neq 0\}$ and so forth.
There are seven different combinations and consequently, seven different phases, which are computed 
and the results collected to generate the complete phase diagrams.

\begin{figure}[ht]
{{\includegraphics [width=0.6\linewidth]{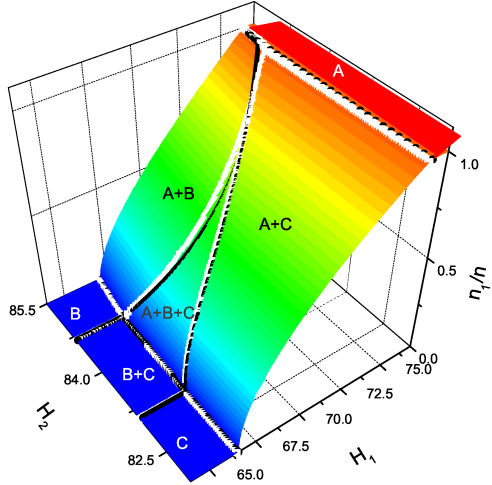}}}
\caption{Phase diagram showing the polarization $n_1/n$ versus the fields $H_1$ and $H_2$ for strong interaction 
with $|c| = 10$ and $n=1$. There are three pure phases: an unpaired phase A, a pairing phase B and 
a trion phase C and four different mixtures of these states.
The black lines are plotted from the analytical results (\ref{H1-H2}) while the white dots
correspond to the numerical solutions of the dressed energy equations (\ref{TBA-F}). The numerical phase transition 
boundaries coincide well with the analytical results.} 
\label{n1strong}
\end{figure}

\begin{figure}[ht]
{{\includegraphics [width=0.6\linewidth]{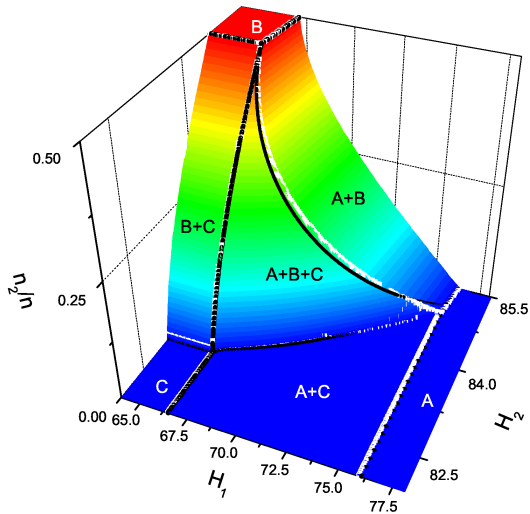}}}
\caption{ Phase diagram showing the polarization $n_2/n$ versus the fields $H_1$ and $H_2$ for strong interaction 
with $|c| = 10$ and $n=1$. The black lines plotted from the analytical results (\ref{H1-H2}) are in good agreement with the 
numerical solutions (white dots) of the dressed energy equations (\ref{TBA-F}). } 
\label{n2strong}
\end{figure}

Figures \ref{n1strong} and \ref{n2strong} show the polarizations $n_1/n$ and $n_2/n$ in terms of the fields $H_1$ and $H_2$.
There are three pure phases: an unpaired phase $A$, a pairing phase $B$ and a trion phase $C$ and four different 
mixtures of these states.
A good agreement is observed between the analytical predictions obtained from equations (\ref {H1-H2}) 
represented by black lines and the numerical solutions obtained by integrating the dressed energy 
equations (\ref{TBA-F}) represented by white dots.

\begin{figure}[t]
{{\includegraphics [width=0.6\linewidth]{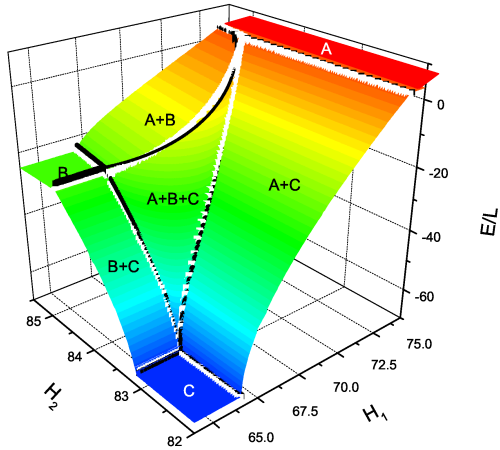}}}
\caption{ Ground state energy {\em vs} Zeeman splitting for strong interaction $|c| = 10$ and $n=1$. 
Good agreement is found between the analytical results (\ref {H1-H2})  represented by black lines and 
the numerical  solutions obtained by integrating the dressed energy equations (\ref{TBA-F}), represented by 
white dots.}\label{fig:E} 
\end{figure}

The ground state energy versus the fields $H_1$ and $H_2$ can be determined from the 
ground state energy (\ref{E}) with the densities $n_1$ and $n_2$ obtained 
from equation (\ref{H1-H2}). 
Figure \ref{fig:E} shows the energy surface for all possible phases 
shown in Figures \ref{n1strong} and \ref{n2strong}.
This figure demonstrates the interplay between different physical ground states.
For  certain values of $H_1$ and $H_2$, a mixture of unpaired fermions, BCS-like 
pairs and trions ($A+B+C$) populates the ground state.

For low temperature, a similar investigation employing the polylog function can also be performed \cite{he}.

\section{Conclusion and perspectives}

We have examined the two- and three-component attractive 1D
Fermi gases with population imbalance via their TBA solution. 
For the two-component model, we reviewed the strong coupling expansion and the identification of quantum phases.
New results for the universal finite temperature corrections were also discussed.
For the three-component model, numerical solution of the dressed energy equations 
confirm that the analytical expressions for the critical fields 
and the resulting phase diagrams at zero temperature 
are highly accurate in the strong coupling regime.
Both models exhibit rich phase diagrams with a variety of quantum phases.
Just as the three distinct phases of the two-component model -- the BCS-like paired, fully polarized and partially 
polarized phases -- have been detected by Liao {\em et al.} \cite{Hulet} in a recent experiment with 
ultracold ${}^{6}$Li atoms in an array of 1D tubes, it is to be hoped that the more exotic phases of the 
three-component model -- including the trion phase -- will be detectable in future experiments.

From the experimental point of view \cite{Hulet}, the array of 1D tubes is created within 2D optical lattices. 
In order to make the lowest transverse mode populated in each tube, the thermal energy $k_BT$ and the 
Fermi energy are required to be much smaller than the transverse confinement energy. 
In this sense, the system is well controllable and practicable only for the strongly attractive regime in the 
quasi-1D trapping.  
Precisely in this regime,  the analytical results may provide direct application to fitting the experimental data, 
such as the density profiles and phase diagram \cite{note}.

Multi-component Fermi gases with more than three species can also be trapped and manipulated. 
For this type of Fermi gas, bound multi-body clusters are expected to appear above certain critical interaction strengths \cite{Luu}.
The thermodynamic properties and phase diagrams of 1D attractive multicomponent Fermi gases can also be investigated
through the solvable models exhibiting $SU(N)$ symmetry \cite{sun}. 
Closed form expressions for the thermodynamics and equation of state of such models will 
provide further insight towards understanding the nature of many-body effects and different pairing states with higher spin symmetry. 
The trapping potentials can be accommodated into the equation of state within the local density approximation. 
It is clear that the Bethe Ansatz will continue to prosper as an essential tool for their description.

\subsection*{Acknowledgments}

A. Foerster and C. C. N. Kuhn  are supported by CNPq (Conselho Nacional de Desenvolvimento
Cient\'{\i}fico e Tecnol\'{o}gico). 
The work of M. T. Batchelor and X.-W. Guan is partially supported by the Australian Research Council.

\end{document}